\documentclass[twocolumn,superscriptaddress,amsmath,amssymb,floatfix] {revtex4}

\pdfoutput=1

\usepackage{graphicx}
\usepackage{dcolumn}
\usepackage{bm}

\graphicspath{{Figures/}}
\setkeys{Gin}{width=\linewidth}
\begin{document}

\title[Optimization of sample-chip design for stub-matched radio-frequency reflectometry measurements]{Optimization of sample-chip design for stub-matched radio-frequency reflectometry measurements}

\author{S.~Hellm\"uller}
\email{hesarah@phys.ethz.ch}
\author{M.~Pikulski}
\author{T.~M\"uller}
\author{B.~K\"ung}
\author{G.~Puebla-Hellmann}
\author{A.~Wallraff}
\author{M.~Beck}
\author{K. Ensslin}
\author{T.~Ihn}
\affiliation{Department of Physics, ETH Zurich, 8093 Zurich, Switzerland\\}

\date{\today}

\begin{abstract}
A radio-frequency (rf) matching circuit with an \textit{in situ} tunable varactor diode used for rf reflectometry measurements in semiconductor nanostructures is investigated and used to optimize the sample-specific chip design. The samples are integrated in a 2-4$\,$GHz stub-matching circuit consisting of a waveguide stub shunted to the terminated coplanar waveguide. Several quantum point contacts fabricated on a GaAs/AlGaAs heterostructure with different chip designs are compared. We show that the change of the reflection coefficient for a fixed change in the quantum point contact conductance can be enhanced by a factor of $3$ compared to conventional designs by a suitable electrode geometry.
\end{abstract}

\maketitle

High-frequency reflectometry measurements in semiconductor nanostructures \cite{schoelkopf1998} allow dispersive state readout detecting the state-dependent ``quantum capacitance'' of a qubit \cite{petersson2010} and time-resolved charge detection measurements \cite{muller2007, reilly2007, cassidy2007, thalakulam2007} using a quantum point contact (QPC) \cite{field1993} or a single-electron transistor \cite{lu2003} as charge detector. We investigate a radio-frequency (rf) matching circuit as an approach to reach higher detection bandwidths in single-shot measurements compared to dc read-out techniques \cite{fujisawa2004, schleser2004, vandersypen2004}. This was motivated by exciting experiments such as the measurement of a single electron spin by spin-to-charge conversion \cite{elzerman2004}, the measurement of full counting statistics \cite{gustavsson2006}, and single-shot readout of a spin qubit \cite{barthel2009}. In contrast to previously demonstrated rf techniques \cite{muller2007, reilly2007, cassidy2007, thalakulam2007}, we use a stub-matching approach, adopted from microwave impedance matching \cite{pozar2005,puebla2012}.

The basic idea of reflectometry measurements is to detect the reflection of a rf-signal depending on the load terminating the waveguide. The reflection coefficient $\Gamma$ is most sensitive to changes of the load impedance $Z_{\mathrm{L}}$ if $Z_{\mathrm{L}}$  is close to the characteristic impedance $Z_\mathrm{0}$ of the waveguide of 50 $\Omega$ \cite{roschier2004}. The reflection coefficient is given by $\Gamma=(Z_{\mathrm{L}}-Z_{\mathrm{0}})/(Z_{\mathrm{L}}+Z_{\mathrm{0}})$. In the presented case the load is a QPC integrated into a rf matching circuit. The latter consists of a waveguide stub shunted to the waveguide terminated by $Z_{\mathrm{L}}$ and is integrated on a printed circuit board (PCB). The shunt-stub circuit is \textit{in situ} tunable which simplifies the matching procedure \cite{mueller2010}. The matching circuit design as well as the sample chip design have a large impact on the measurement performance. In this paper we focus on the optimization of the chip design. Previous transmission measurements in GaAs/AlGaAs heterostructures point to the assumption that these results may also be important for matching frequencies  lower than the  2-4$\,$GHz investigated in this paper \cite{c1}. The paper is written within the scope of time-resolved charge detection measurements. The conclusions for the design of the sample chip  and the PCB, however, are more generally valid for reflectometry measurements.

\begin{figure}[!hb]
\begin{center}
\includegraphics [width=0.47\textwidth ]{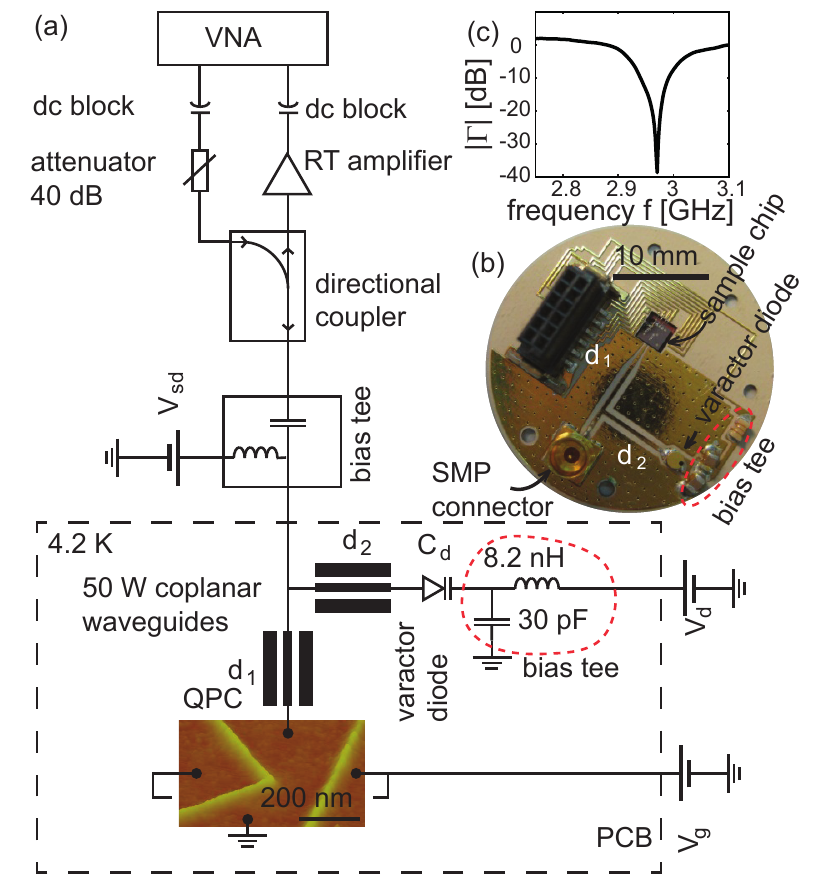}
\end{center}
\caption{(a) Experimental setup used to perform the measurements and designed for carrier frequencies of $2-4 \,$GHz. (b) Photograph of the PCB with the two stubs terminated by a varactor diode and the QPC, respectively. Two capacitors and one inductor placed after the diode act as on-chip bias tee. (c) Absolute value of the reflection coefficient $\left| \Gamma \right|$ as a function of the carrier frequency $f$ for a QPC conductance of $g=3 \times 2e^2/h$.}
\label{fig1}
\end{figure}

In the following, we discuss the setup sketched in Fig.$\,$\ref{fig1}(a). A rf carrier signal is generated by a vector network analyzer (VNA) and applied to the sample via a directional coupler. Before reaching the sample, a dc component is added to the signal by using a bias tee. At the PCB, containing matching circuit and sample, the rf signal is partly reflected and afterwards amplified by 60$\,$dB using a 2-4$\,$GHz room temperature amplifier before reaching the input port of the VNA. Dc blocks are used to avoid ground loops. All components except for the PCB are placed at room temperature for ease of calibration. The calibration is done in 3 steps using commercial calibration kits\cite{c2}. 

A schematic of the PCB is shown in the dashed box in Fig.$\,$\ref{fig1}(a) and on the photograph in Fig.$\,$\ref{fig1}(b). The substrate of the PCB is AD1000 (Arlon), with a dielectric constant of 10 and a roughly 17$\,\mu$m thick copper layer. A thin gold coating prevents oxidation. The transmission lines on the PCB, coplanar waveguides with ground plane (CPWG), form the matching circuit and have a loss of about 3.88$\,$dB/m at 3$\,$GHz. The sample chip including the QPC (shown in the atomic force microscope (AFM) scan in Fig.$\,$\ref{fig1}(a)) is glued into a \mbox{$3 \times 3$ mm$^2$} recess at the end of the CPWG and connected with bond wires. 

Impedance matching is achieved by a CPWG stub shunted to the 50$\,\Omega$ CPWG at a distance $d_\mathrm{1}$ of about $\lambda$/4 from the QPC load, where $\lambda$ is the wavelength of the roughly 3$\,$GHz carrier signal. The stub has a length $d_\mathrm{2}$ of about $\lambda$/4 and is terminated by a varactor diode. The capacitance $C_{\mathrm{d}}$ of the diode is \textit{in situ} tunable by applying a dc voltage $V_{\mathrm{d}}$  \cite{mueller2010}. A bias tee, consisting of two on-chip capacitors of 15 pF each and of an inductor of 8.2 nH, is placed on the PCB close to the diode. This provides a rf ground for the rf signal without affecting the dc voltage $V_{\mathrm{d}}$ tuning the diode's capacitance.  The influence of the bias tee on the rf circuit performance is irrelevant as the diode capacitance $C_{\mathrm{d}}$ of $0.3-1.1\,$pF is small enough compared to the capacitance of the on-chip bias tee. The total impedance $Z_{\mathrm{L}}$ of the stub-matching circuit with integrated QPC and diode is given by the inverse of the sum of the two admittances - the one of the CPWG part from the load to the stub including the QPC $Y_{d_1}$ and the one of the stub including the diode $Y_{d_2}$
\begin{eqnarray}
Z_{\mathrm{L}} = Y_{\mathrm{L}}^{-1} = \left(Y_{d_1}+Y_{d_2}\right)^{-1}
=\nonumber\\
\left( 
Y_{\mathrm{0}} \frac{Y_{\mathrm{QPC}}+Y_{\mathrm{0}}\,\mathrm{tanh}((\alpha + i2\pi/\lambda) d_{1})}{Y_{\mathrm{0}}+Y_{\mathrm{QPC}}\,\mathrm{tanh}((\alpha + i2\pi/\lambda) d_{1})} 
+\right. \nonumber\\
\left.
Y_{\mathrm{0}} \frac{Y_{\mathrm{d}}+Y_{\mathrm{0}}\,\mathrm{tanh}((\alpha + i2\pi/\lambda) d_{2})}{Y_{\mathrm{0}}+Y_{\mathrm{d}}\,\mathrm{tanh}((\alpha + i2\pi/\lambda) d_{2})}
\right)^{-1},\nonumber
\end{eqnarray}
where $Y_{\mathrm{d}}$ is the admittance of the diode and $\alpha$ is the attenuation constant of the CPWG. The basic idea to match $Z_\mathrm{L}$ to $Z_\mathrm{0}=(Y_0)^{-1}=50\,\Omega$ is to choose $d_\mathrm{1}$ such that for a certain frequency $f$ the admittance of the load terminating the CPWG, in our case $Y_{\mathrm{QPC}}$, is tranformed to $Y_{d_1}=Y_{\mathrm{0}} + jB$ at the distance $d_\mathrm{1}$ from the load. The susceptance $B$ depends on $Y_\mathrm{QPC}$. The length $d_\mathrm{2}$ of the shunted stub is then designed such that the admittance of the stub is equal to $Y_{d_2}=-jB$ and hence the total admittance $Y_{\mathrm{L}}$ is exactly the characteristic admittance $Y_\mathrm{0}$. 

For fixed $d_\mathrm{1} = 8.23\,\mathrm{mm}$ and $d_\mathrm{2} = 8.47\,\mathrm{mm}$, as in our case, the carrier frequency $f$ and the capacitance of the varactor diode $C_{\mathrm{d}}$ are adjusted to achieve impedance matching ($Z_{\mathrm{L}}=Z_0$ and hence $\left|\Gamma\right|$=0). For distinct QPC conductance values, the reflection coefficient is measured as a function of the diode voltage $V_{\mathrm{d}}$ and the carrier frequency $f$, and pairs ($f_\mathrm{m}$,$V_{\mathrm{d}_\mathrm{m}}$) can be extracted for which $Z_\mathrm{L}$ is closest to 50$\,$$\Omega$. In Fig.$\,$\ref{fig1}(c) the absolute value of the reflection coefficient $\left|\Gamma\right|$ versus frequency $f$ for a fixed diode voltage is shown. The PCB is designed for a matching frequency close to 3 GHz, and we measure a full width at half maximum of around 100 MHz, which puts an upper limit on the setup bandwidth.

The samples are fabricated using photolithography and AFM lithography on a GaAs/AlGaAs heterostructure with a two-dimensional electron gas (2DEG) 34$\,$nm below the surface. The 2DEG has a mobility of \mbox{33$\,$m$^2$/Vs} and an electron density of \mbox{$4.8\times 10^{15}\,$m$^{-2}$} at a temperature of 4$\,$K. 

\begin{figure}[!ht]
\begin{center}
\includegraphics [width=0.5\textwidth ]{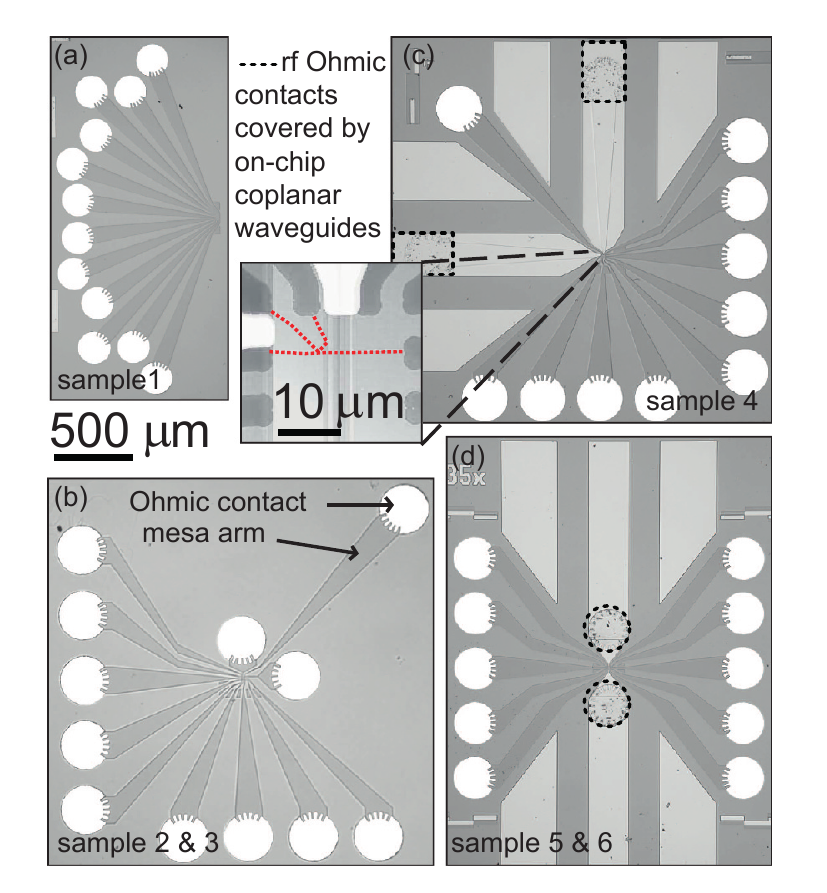}
\end{center}
\caption{Photographs of samples with the same design as (a) sample 1, (b) sample 2 and 3, (c) sample 4, (d) sample 5 and 6. The white circles are conventional Ohmic contacts and the darkest grey areas are the mesa arms where the 2DEG remains after etching. In (c) and (d) the lightest grey areas are the coplanar waveguides where the thick dotted lines indicate additional Ohmic contacts. In the inset of (c), an AFM scan of the central mesa is shown and the fine red dotted line marks the AMF lithography oxide lines.}
\label{fig2}
\end{figure}

We compare 6 samples with 4 different designs shown in Fig.$\,$\ref{fig2}(a-d). The mesas (darkest grey areas) are the conducting plateaus that remain after etching the GaAs/AlGaAs heterostructure. Dc voltages are applied to the quantum structure via Ohmic contacts (white circles), including special rf Ohmic contacts (areas within dotted black lines) placed at different distances to the center mesa area, where a QPC is located. The rf Ohmic contacts in Fig.$\,$\ref{fig2}(c,d) are covered by 200$\,$nm thick gold electrodes (light grey areas) forming CPWGs to optimize the rf performance. The CPWGs loss of about 30$\,$dB/m is higher than the loss on the PCB but it is still not comparable to the resistive damping in the 2DEG \cite{burke2000}. The inset of Fig.$\,$\ref{fig2}(c) shows the AFM scan of \mbox{sample 4} including the AFM lines (red dotted line) forming the QPC. 

\begin{figure}[t!]
\begin{center}
\includegraphics [width=0.5\textwidth ]{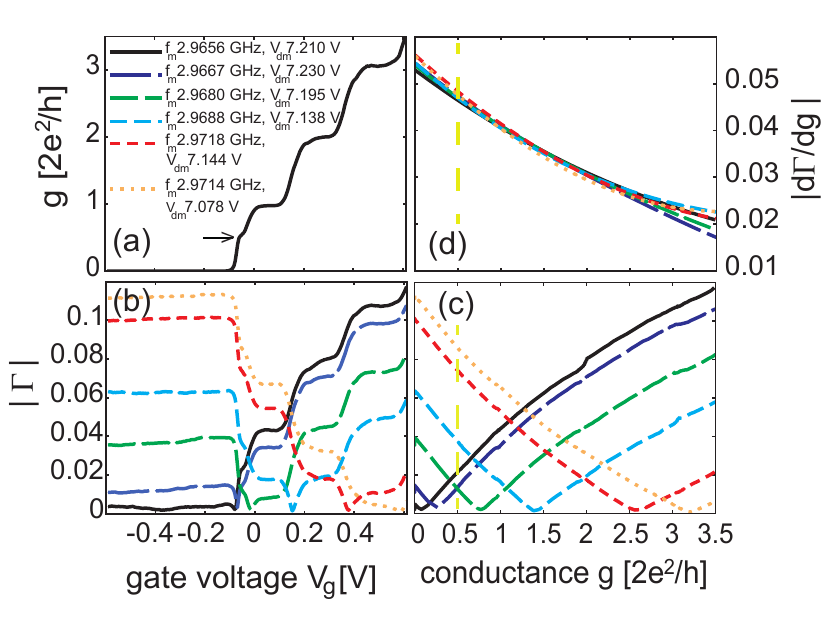}
\end{center}
\caption{(a) The dc conductance of the QPC as a function of gate voltage $V$$_{\mathrm{g}}$. (b) The absolute value of the reflection coefficient $\left|\Gamma\right|$ depending on the gate voltage for different carrier frequencies $f_{\mathrm{m}}$ and different voltages $V_{\mathrm{d}_\mathrm{m}}$ applied to the diode. (c) $\left|\Gamma\right|$ as a function of the conductance obtained by combining the data of (a) and (b). (d) The derivative of $\Gamma$ as a function of $g$. The dashed yellow line indicates the $g$ value for which $\left|\mathrm{d}\Gamma/\mathrm{d}g\right|$ is the figure of merit for the sample.}
\label{fig3}
\end{figure}

\begin{table}[b]
\caption{Characteristics of samples 1 to 6 including distance from the rf Ohmic contact to the QPC (long ($\ge$$\,$785$\,\mu$m) or short ($\le$$\,$115$\,\mu$m)), on-chip coplanar waveguides (existent/inexistent) and wedge (W, samples \mbox{1-4a}, 5, 6a) or ribbon (R, samples 4b, 6b) bonds to connect the samples to the PCB.}
\begin{tabular*}{0.5\textwidth}{@{\extracolsep{\fill}}lllll}
									\hline
									\hline
									& distance&on-chip	& \\	
								  sample no.& to QPC&CPWG 	 &bonds	& Fig.$\,$\ref{fig2} \\				
									\hline
									1 &$\sim$$\,1000\,\mu$m& no&W& (a)\\
									2,3&$\sim$$\,75\,\mu$m&no&W& (b)\\
									4a,4b&$\sim$$\,785\,\mu$m&yes&W,R& (c)\\
									5&$\sim$$\,115\,\mu$m&yes&W& (d)\\
									6a,6b&$\sim$$\,110\,\mu$m&yes&W,R& (d)\\		
									\hline
									\hline
									\label{tab1}
\end{tabular*}
\end{table}

In a charge detection measurement, charging a quantum dot (QD) by one electron causes a certain change in the QPC conductance. The maximum QPC conductance change due to one additional electron on the QD $\left|\mathrm{d}g/\mathrm{d}q\right|$  is a quantity that varies from one nanostructure to the next and is not a characteristic of the microscale chip design but depends on the nanostructure design. To measure the performance of the chip design, we therefore consider the change of the reflection coefficient for a given change in the QPC conductance $\left|\mathrm{d}\Gamma/\mathrm{d}g\right|$ by varying the voltage $V_\mathrm{g}$ on a gate electrode next to the QPC \cite{c3}. 
The figure of merit for an optimized design is the slope $\left|\mathrm{d}\Gamma/\mathrm{d}g\right|$ of the curves evaluated at the conductance $g$ where $\left|\mathrm{d}\Gamma/\mathrm{d}g\right|$$\left|\mathrm{d}g/\mathrm{d}q\right|$ has its maximum and this is usually close to $0.5\times2e^2$/$h$ as $\left|\mathrm{d}g/\mathrm{d}q\right|$ has a maximum there (arrow in Fig.$\,$\ref{fig3}(a)). Therefore it is a good measure to compare $\left|\mathrm{d}\Gamma/\mathrm{d}g\right|$ at $g$ equal to $0.5\times2e^2$/$h$ for different sample designs.

Fig.$\,$\ref{fig3}(a) shows the quantized conductance of the QPC as a function of gate voltage $V_{\mathrm{g}}$. A gate voltage independent contact resistance is subtracted from the measured resistance. These contact resistances were extracted to be between 1.2$\,$-$\,$4$\,$k$\Omega$ for all samples except for sample 4a (9$\,$k$\Omega$) and 4b (15$\,$k$\Omega$). 

We will now direct our attention to the rf measurements. The carrier signal power reaching the PCB is -65$\,$dBm. The absolute value of the reflection coefficient  $\left|\Gamma\right|$ is plotted as a function of QPC gate voltage for selected pairs ($f_\mathrm{m}$,$V_{\mathrm{d}_\mathrm{m}}$) in Fig.$\,$\ref{fig3}(b), where $f_\mathrm{m}$ is the matching frequency and $V_{\mathrm{d}_\mathrm{m}}$ is the diode voltage applied to achieve matching. The plateaus due to the quantized conductance of the QPC are clearly observable. By combining the data shown in Fig.$\,$\ref{fig3}(a) and \ref{fig3}(b), $\left|\Gamma\right|$ can be plotted as a function of conductance $g$ (see Fig.$\,$\ref{fig3}(c)). In Fig.$\,$\ref{fig3}(d), the slope $\left|\mathrm{d}\Gamma/\mathrm{d}g\right|$ as a function of $g$ is plotted \cite{c4}. It does not change significantly for curves with different matching points (conductance values for which $\left|\Gamma\right|$=0 holds, corresponding to a pair ($f_\mathrm{m}$,$V_{\mathrm{d}_\mathrm{m}}$)) from 0 to $3\times2e^2$/h (see Fig.$\,$\ref{fig3}(d)).
But if the matching point is moved to even higher conductance values, a decrease in $\left|\mathrm{d}\Gamma/\mathrm{d}g\right|$ is expected for any $g$ close to the pinch-off of the QPC (not shown). 

\begin{figure}[!ht]
\begin{center}
\includegraphics [width=0.5\textwidth ]{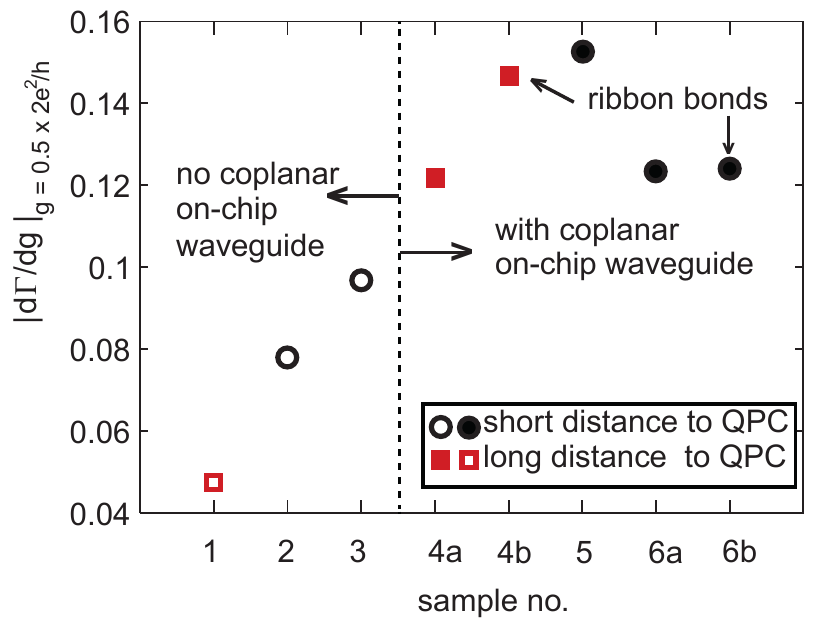}
\end{center}
\caption{$\left|\mathrm{d}\Gamma/\mathrm{d}g\right|$$_{g\,=\,0.5\times2\mathrm{e}^2\mathrm{/h}}$ for the 6 investigated samples. On-chip extensions of the coplanar waveguides (full squares/circles) increase the performance of the samples significantly. The distance from the Ohmic contact to the QPC (long: red squares, short: black circles) only plays a role for the samples without on-chip CPWG (empty squares/circles).}
\label{fig4}
\end{figure}

Tab.$\,$\ref{tab1} gives an overview of the main parameters of the 6 devices with the designs shown in Fig.$\,$\ref{fig2}. One characteristic is the distance between the QPC and the Ohmic contact through which the rf signal is applied. Long distances are $\ge$$\,$785$\,\mu$m and short distances are  $\le$$\,$115$\,\mu$m. Half of the samples have a 50$\,$$\Omega$ on-chip CPWG which ends $\sim$$\,$10$\,$$\mu$m before the QPC. Another criterion is the bonding. All samples are equipped with wedge bond wires (W, diameter \mbox{30 $\mu$m}) and sample 4 and 6 additionally with ribbon bond wires (R, \mbox{$100 \times 12$ $\mu$m} cross-sectional area, made out of Au). 

The results are presented in Fig.$\,$\ref{fig4} where $\left|\mathrm{d}\Gamma/\mathrm{d}g\right|$$_{g=0.5 \times 2e^2/h}$ is plotted for each sample for one representative pair ($f_\mathrm{m}$,$V_{\mathrm{d}\, \mathrm{m}}$). It is clearly visible that all samples with on-chip CPWG (full squares/circles) perform better than the ones without on-chip CPWG (empty squares/circles), independent of the distance from the Ohmic contact to the QPC. A factor of 3 improvement from the standard sample (no. 1, distance $\ge$$\,$785$\,\mu$m, no on-chip CPWG) to the best sample (no. 5, distance $\le$$\,$115$\,\mu$m, with on-chip CPWG) was achieved. For the samples without on-chip CPWG, the distance (long: red squares, short: black circles) over which the rf signal has to pass through the 2DEG before reaching the QPC seems to be a relevant criterion. One more sample supporting this result was measured, but as it was not possible to close the QPC to $0.5\times2e^2$/h (due to leakage currents) the data is not shown here. The use of ribbon bonds for sample 4 shows an improvement compared to wedge bonds. All 6 curves measured at different matching points (same procedure as shown in Fig.$\,$\ref{fig3}) for sample 4b have a larger $\left|\mathrm{d}\Gamma/\mathrm{d}g\right|$ compared to all curves of sample 4a. However, for sample 6 no conclusion is possible as the on-chip CPWG was partly damaged during the bonding procedure. To check the reproducibility, sample 2 and 6b were cooled down again, showing only small deviations. Furthermore, by moving to a different matching point, it was possible to improve the sensitivity by up to 11.5\% at points not measured in the first cooldown.

In conclusion, we have presented an \textit{in situ} tunable shunt stub matching circuit and observed that the sample chip design can be improved significantly leading to up to a factor of 3 improvement in $\left|\mathrm{d}\Gamma/\mathrm{d}g\right|$$_{g=0.5 \times 2e^2/h}$ by integrating coplanar waveguides into the design. A smaller improvement is achieved by simply reducing the distance between Ohmic contact and QPC. These designs can be adapted to other rf-matching circuits as well. The design of the matching circuit integrated into the PCB can be adjusted to higher resonance frequencies by decreasing the stub lengths accordingly. The tunability of the matching circuit facilitates impedance matching and the measurement of a variety of different samples in the future such as InAs nanowire or graphene quantum structures \cite{choi2012, guettinger2011} which are promising for an increased coupling between detector and nanostructure to measure.
\newline

We thank P. Leek, P. Studerus, C. Barengo, and H. Rusterholz for technical discussions and contributions. This research was supported by the Swiss National Science Foundation through the National Centre of Competence in Research Quantum Science and Technology

%


\begin{thebibliography}{10}
\bibitem{c1}%
  \BibitemOpen
  \bibfield{author}{%
  \bibinfo {author} {\bibfnamefont{}\bibnamefont{These transmission experiments compare coplanar waveguides where the waveguide is formed either by the not etched two-dimensional electron gas (2DEG) parts in a GaAs/AlGaAs heterostructure or by gold electrodes evaporated on top of a GaAs/AlGaAs heterostructure. The additional loss in the 2DEG was roughly 10 dB/mm for frequencies down to a few hundred MHz}}\bibinfo
  {author}}%
  \BibitemShut{NoStop}%
\bibitem{c2}%
  \BibitemOpen
  \bibfield{author}{%
  \bibinfo {author} {\bibfnamefont{}\bibnamefont{The reference plane is moved from the two cables connected to the VNA to the last SMA connector at RT and finally to the SMP connector to which the PCB is attached}}\bibinfo
  {author}}%
  \BibitemShut{NoStop}%
\bibitem{c3}%
  \BibitemOpen
  \bibfield{author}{%
  \bibinfo {author} {\bibfnamefont{}\bibnamefont{We assume that in future experiments amplitude and phase sensitive measurements can be performed}}\bibinfo
  {author}}%
  \BibitemShut{NoStop}%
\bibitem{c4}%
  \BibitemOpen
  \bibfield{author}{%
  \bibinfo {author} {\bibfnamefont{}\bibnamefont{The data is fitted with a cubic polynomial. The derivative of the fit is shown. The numerical derivation after averaging the data leads to the same but more noisy result}}\bibinfo
  {author}}%
  \BibitemShut{NoStop}%
\end{thebibliography}
\end{document}